\documentclass{article}
\usepackage{spconf,amsmath,amssymb,graphicx}
\usepackage{tabularx}
\usepackage{booktabs}

\usepackage{xspace}
\usepackage{xcolor}

\definecolor{AlexColor}{rgb}{1.0,0.0,0.0}
\definecolor{PeterColor}{rgb}{0.0,0.0,0.8}
\definecolor{VamsiColor}{rgb}{0.8,0.8,0.0}
\definecolor{DejanColor}{rgb}{0.0,0.0,1.0}

\usepackage{fancyhdr}% http://ctan.org/pkg/fancyhdr
\makeatletter
\DeclareRobustCommand\onedot{\futurelet\@let@token\@onedot}
\def\@onedot{\ifx\@let@token.\else.\null\fi\xspace}

\def\ie{\emph{i.e}\onedot}

\makeatother

\title{Deep Impulse Responses: Estimating and Parameterizing Filters \\with Deep Networks}
\name{Alexander Richard, Peter Dodds, Vamsi Krishna Ithapu}
\address{Reality Labs Research}
\begin{document}
\ninept

\thispagestyle{fancy}
\maketitle
\begin{abstract}
%Estimation of impulse responses has been a core problem in signal processing for decades.
%Yet, accurately measuring and estimating impulse responses in 3D scenes - a problem of significant interest for AR/VR and computer games - is still challenging due to complex spatio-temporal signal characteristics, noise, and uncontrolled recording setups in real-world scenarios.
Impulse response estimation in high noise and in-the-wild settings, with minimal control of the underlying data distributions, is a challenging problem.
We propose a novel framework for parameterizing and estimating impulse responses based on recent advances in neural representation learning.
Our framework is driven by a carefully designed neural network that jointly estimates the impulse response 
and the (apriori unknown) spectral noise characteristics of an observed signal given the source signal.
We demonstrate robustness in estimation, even under low signal-to-noise ratios, 
and show strong results when learning from spatio-temporal real-world speech data.
Our framework provides a natural way to interpolate impulse responses on a spatial grid, 
while also allowing for efficiently compressing and storing them for real-time rendering applications in augmented and virtual reality. 
%that have spatial relationships, and can parameterize families of impulse responses in a memory efficient way, 
%making it particularly interesting for on-device applications in AR and VR.
\end{abstract}
%

%%%%%%%%%%%%%%%%%%%%%%%%%%%%%%%%%%%%%%%%%%%%%%%%%%%%%%%%%%%%%%%%%%%%%%%%%%%%%%%%%%%%%%%%%%%%%%%%%%%%%%%%%%%%%%%%%%%%%%%%%%%%%%%%%%%%%%%%%%%%%%%%%%%%%%%%%%%%
%%% INTRODUCTION
%%%%%%%%%%%%%%%%%%%%%%%%%%%%%%%%%%%%%%%%%%%%%%%%%%%%%%%%%%%%%%%%%%%%%%%%%%%%%%%%%%%%%%%%%%%%%%%%%%%%%%%%%%%%%%%%%%%%%%%%%%%%%%%%%%%%%%%%%%%%%%%%%%%%%%%%%%%%

\section{Introduction}
\label{sec:intro}

Robust estimation of impulse responses (IRs) is vital for a variety of audio and speech signal processing applications, 
including scene and room acoustics modeling, source localization, and audio spatialization. 
Accurately characterizing room and head-related impulse responses (RIR and HRIR) 
is required for achieving immersion and realism, 
while enabling audio and sound pesonalization in augmented and virtual reality applications.

Typically these IRs are spatio-temporal, \ie, each spatial location in $3$D corresponds to one short time-domain signal. 
%An alternative to estimating full IR is to instead estimate certain sufficient statistics (e.g., $T_{60}$~\cite{?} or Direct-to-Reverberant Ratio~\cite{?} with RIRs), or utilize a minimum-phase approximation and estimate the frequency domain transfer function (e.g., HRTFs)~\cite{?}. 
%Nevertheless, both these approaches are approximations and ideally modeling the full IR is desired. 
Measuring IRs for every spatial location in a $3$D scene is infeasible in many cases due to the cost and complexity of the setup involved and the setups themselves are typically not portable. %, thereby requiring one to recreate them with every room (scene) and/or user involved. 
Hence, reliable estimation and prediction of such IRs using computational models has received significant 
attention in the audio and speech signal processing community \cite{lin2006bayesian, szoke2019building, xie2007head}. 
IR estimation in linear time invariant (LTI) systems typically corresponds to solving an inverse problem \cite{dinuzzo2015kernels} and traditional approaches would require clean data or assume additive noise with known spectral noise characteristics. 
Neither of these are available for in-the-wild scenarios. 
While optimal or adaptive filtering techniques have been proposed, they require \textit{apriori} knowledge or some estimate of the underlying signal statistics~\cite{wiener1949filter,widrow1975lms,liu2010adaptivefiltering}.

\begin{figure}[tb]
    \centering
    \includegraphics[scale=0.3]{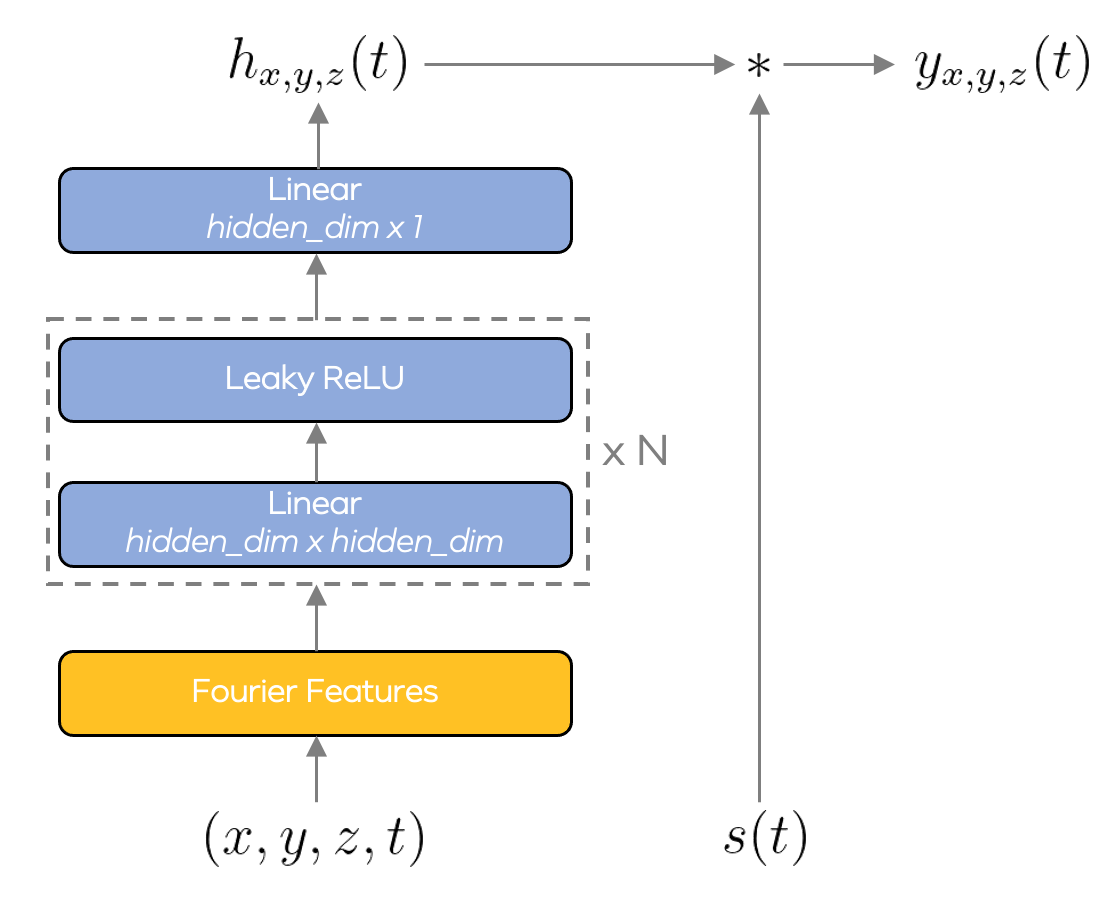}
    \vspace{-0.3cm}
    \caption{The proposed IR-MLP.
             Impulse responses are predicted from spatio-temporal coordinates using an MLP.
             The final result is obtained by convolution of the IR with the source signal.
             %Note that the process is fully differentiable.
             }
    \label{fig:model}
    \vspace{-0.5cm}
\end{figure}

%Alternatively, compressive sensing and signal reconstruction techniques have also been explored for IR estimation \cite{?}. 
%They rely on using a sparse set of reliable measurements on a fixed spatial grid, and a dense interpolation for recovering IRs for the rest.
Besides robust estimation of IRs in noisy in-the-wild scenarios, interpolation of estimated IRs poses a major challenge.
As it is infeasible to measure IRs at every spatial position, bilinear or barycentric interpolation are often utilized to approximate IRs at unmeasured positions~\cite{antonello2017room, ramos2013parametric, gamper2013hrtfinterp}.
However, such interpolation methods assume that IRs undergo strictly linear transformations as the spatial locations change, which is not an accurate representation of reality.
%struggle to generate high quality interpolated filters particularly from sparse measurements.
More sophisticated IR interpolation techniques that are domain-specific have been proposed, 
however they suffer from generalization to arbitrary measurement grids and application domains \cite{das2021room}.

In this work, we mitigate some of these drawbacks in IR estimation and interpolation by leveraging recent advances from neural representation learning~\cite{sitzmann2019siren,tancik2020fourierfeatures,mildenhall2020nerf}.
We propose a simple model driven by multi-layer perceptrons (MLPs) and a novel loss function 
for training the MLPs to jointly estimate the IR and the unknown noise characteristics.
Deep networks have recently shown to be effective for learning time domain representations \cite{vandenoord2016wavenet,fu2017raw,luo2018tasnet,donahue2019wavegan} and MLPs are one widely studied architectural family of deep networks that are successful in acoustic modeling for speech recognition~\cite{richard2014mnsgd}, audio equalization~\cite{pepe2020designing}, and speech enhancement~\cite{nossier2020comparative}. 
MLPs have not only been proven to be able to model high frequency signals~\cite{tancik2020fourierfeatures} but also are excellent interpolation machines, providing a natural mechanism to generate IRs at unmeasured positions.
While some recent existing works estimate IRs with neural networks~\cite{richard2021binaural, gebru2021implicit, steinmetz2021filtered, ratnarajah2020ir}, 
they rely mainly on domain-specific architectures carefully designed to capture some underlying apriori data attributes~\cite{gamper2018blind, bryan2020impulse, yu2020room}, 
and do not always yield strong results~\cite{steinmetz2021filtered}.
Our framework, in contrast, is generic, without domain-specific assumptions, and applicable to any problem where IR estimation is required.
%
%Lastly, efficient storage of IRs is vital for real-time applications in AR and VR, especially on small scale form-factors like glasses. 
%We show that the proposed MLP-based IR parameterization can reduce the memory consumption by several orders of magnitude at minimal loss of precision.
%This not only allows for adaptive modification of room IRs, for example, but also allows for 
%cheap personalization of HRIRs based on user specific information.
%Overall the proposed learning framework (a) estimates IRs in-the-wild with no assumptions on noise structure, 
%(b) allows for efficient interpolation of IRs, and (c) parameterizes IRs for efficient compression and storage.
%
In summary, our contributions are:\\
\textbf{Impulse response estimation.} We propose an effective and robust method to estimate impulse responses in a highly noisy setting where traditional approaches fail or perform worse.\\
\textbf{Efficient parameterization of impulse responses.} We demonstrate that our approach can store impulse responses with an extremely high compression factor and therefore an extremely low memory footprint, which is a key property for on-device applications.\\
\textbf{Native interpolation of impulse responses.} We show that the parameterization with a neural network is not only highly memory efficient but also allows to interpolate impulse responses at unseen positions more accurately than traditional interpolation techniques.

\section{Model Formulation} \label{sec:model}

Consider an LTI system characterized by spatio-temporal IRs, where $(x, y, z)$ correspond to the $3$D position of source (or receiver).
We restrict ourselves to $3$D spatial positions in this work for clarity and simplicity.
Extension to higher dimensional cases is straightforward.
If $ s(t) $, $ h(t) $ and $y(t)$ are the source, unknown IR, and the observed (target) signal respectively, then we have
\begin{align} \label{eq:lti}
    y_{x,y,z}(t) = s(t) * h_{x,y,z}(t).
\end{align}
%
%Note that the proposal in this work can be applied to filters parameters by any regular grids (e.g., conditioned on some inputs).
%
%where $ s(t) $ and $ h(t) $ are the source and unknown impulse response respectively.
%We are interested in spatio-temrapal IRs parameterized by the 3D position $ (x, y, z) $ of source (or receiver), 
%In many applications, instead of a single impulse response, a family of impulse responses needs to be estimated.
%Prominent examples are room impulse responses or head-related transfer functions.
%In those cases, the filter depends on the 
%We will stick to the example of families of filters that are conditioned on spatial positions throughout this work but families of filters characterized by any other conditional input can be handled in the same way.
%
Estimating such families of filters means to learn a function
\begin{align}
    \mathcal{F}: (x,y,z,t) \mapsto h_{x,y,z}(t)
\end{align}
that maps from positional inputs $ (x,y,z) $ and the temporal index $ t $ of the filter to the $ t $-th sample of the filter.
We parameterize $ \mathcal{F} $ with an MLP that consumes, as input, the spatial coordinates $ x,y,z $ and the temporal index $ t $ and predicts, as output, the $ t $-th sample of the finite impulse response $ h_{x,y,z}(t) $, see Figure~\ref{fig:model}.
As pointed out in~\cite{tancik2020fourierfeatures}, MLPs typically struggle to learn high frequency functions in low dimensional domains.
We therefore adopt a popular solution used in neural rendering~\cite{mildenhall2020nerf} and compute Fourier features
\begin{align}
    \gamma(p) = \big\{ \sin(2^\ell \pi p), \cos(2^\ell \pi p) ; 0 \leq \ell < L \big\}
    \label{eq:fourier_features}
\end{align}
of the low dimensional input $ p := (x,y,z,t) $ before feeding them into the MLP.
%
%We use $ L=10 $ in all our models.
%
The output of the LTI from Equation~\eqref{eq:lti} is then obtained by convolution of the input signal $ s(t) $ with the predicted FIR $ h_{x,y,z}(t) $.
%
%Since the convolution operation is differentiable, the MLP can be trained by computing a loss directly between the measured and predicted waveforms.
We use an $ \ell_2 $-loss for training the MLP:
\begin{align}
    \mathcal{L} = \sum_t \|\hat y_{x,y,z}(t) - y_{x,z,y}(t)\|^2,
\end{align}
where $ \hat y_{x,y,z}(t) $ denotes the (measured) ground truth signal and $ y_{x,y,z}(t) $ is the signal obtained by convolution of the MLP output with the input signal.

%Note that we restrict ourselves to IRs which depend on spatial positions in this work for clarity and simplicity.
%
%In fact, the approach can be applied to estimate IRs conditioned on any manifold rather than just on positions in a 3D Euclidean space.

%%%%%%%%%%%%%%%%%%%%%%%%%%%%%%%%%%%%%%%%%%%%%%%%%%%%%%%%%%%%%%%%%%%%%%%%%%%%%%%%%%%%%%%%%%%%%%%%%%%%%%%%%%%%%%%%%%%%%%%%%%%%%%%%%%%%%%%%%%%%%%%%%%%%%%%%%%%%
%%% NOISE ROBUST LEARNING
%%%%%%%%%%%%%%%%%%%%%%%%%%%%%%%%%%%%%%%%%%%%%%%%%%%%%%%%%%%%%%%%%%%%%%%%%%%%%%%%%%%%%%%%%%%%%%%%%%%%%%%%%%%%%%%%%%%%%%%%%%%%%%%%%%%%%%%%%%%%%%%%%%%%%%%%%%%%

\section{Noise-Robust Learning}
\label{sec:learning}

\begin{figure}
    \centering
    \includegraphics[scale=0.27]{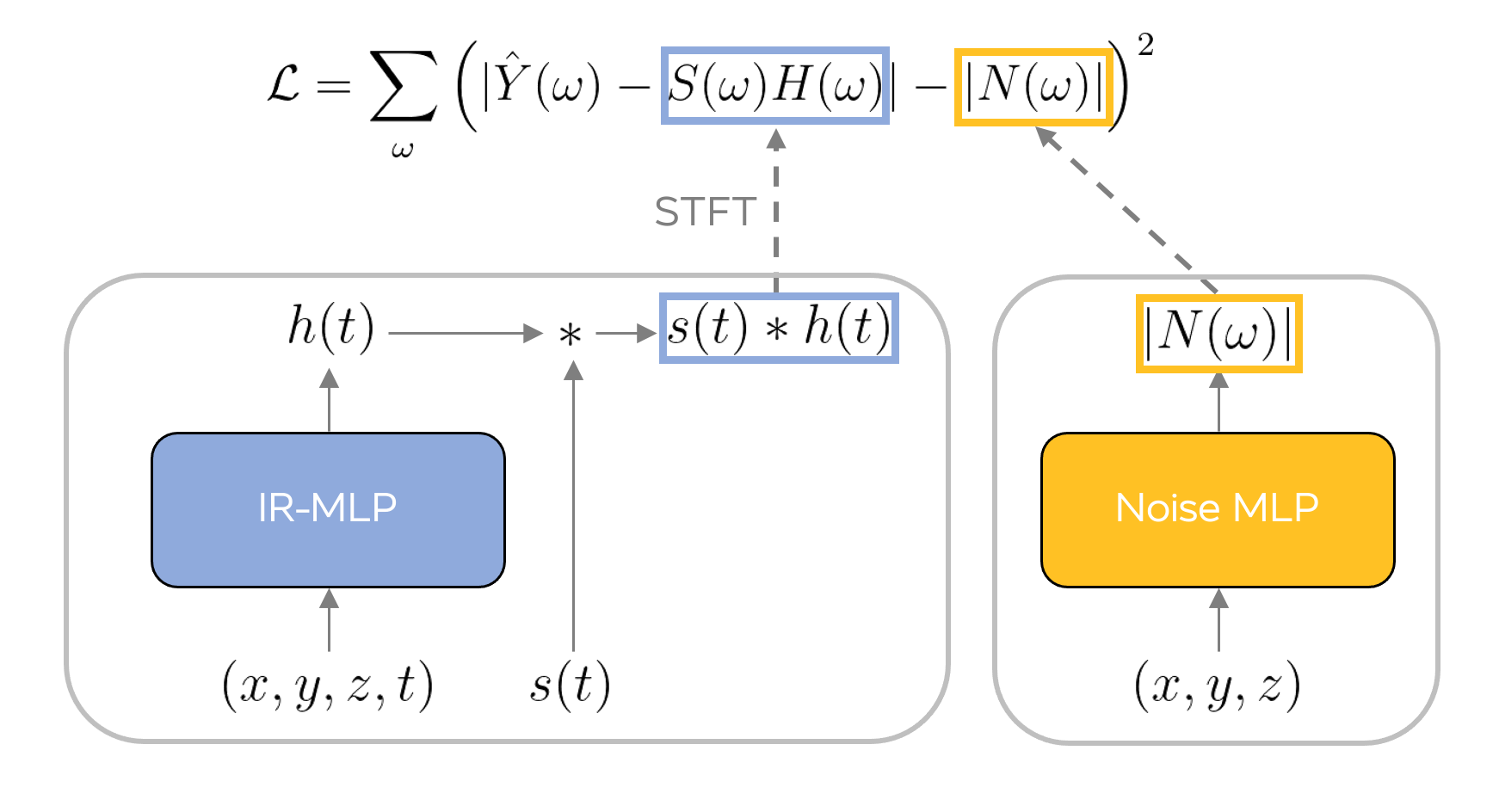}
    \vspace{-0.3cm}
    \caption{In addition to the IR-MLP, a Noise MLP predicts the spectral characteristics of additive noise.
             A noise-robust loss function is then optimized to jointly learn the IRs and noise spectral characteristcs.}
    \label{fig:loss}
\end{figure}

\begin{figure*}[tb]
    \centering
    \includegraphics[scale=0.8]{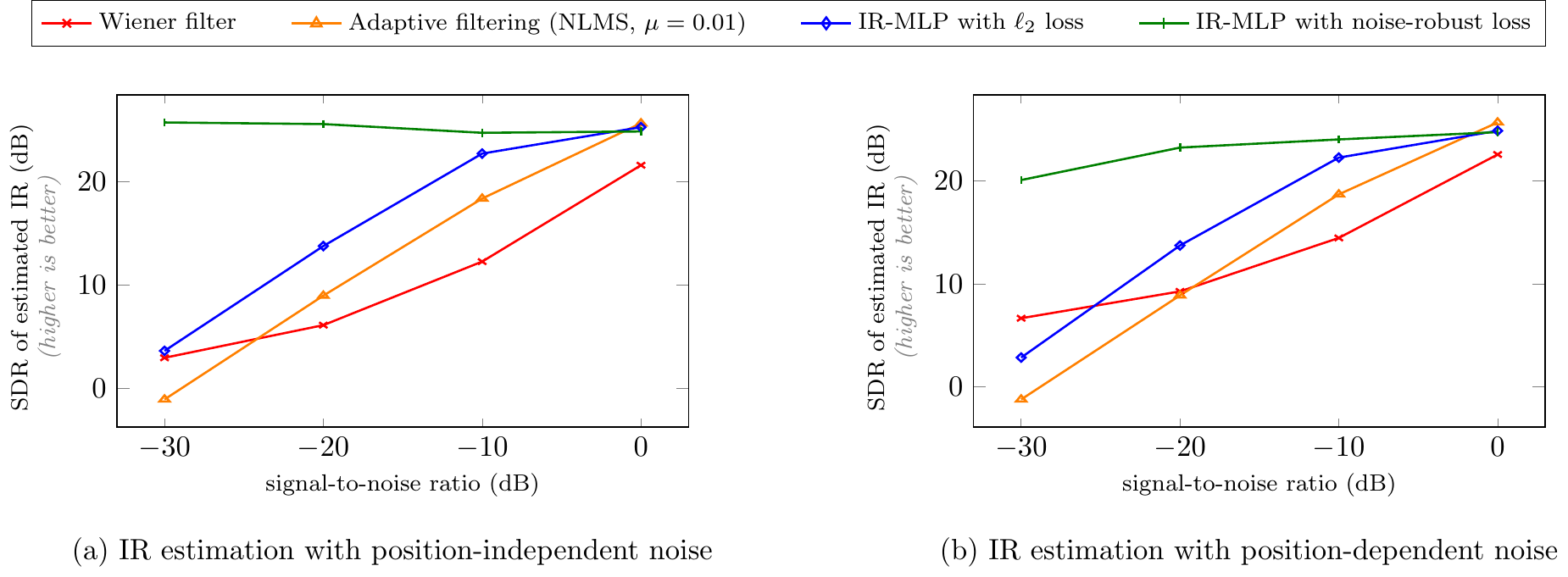}
    \vspace{-0.3cm}
    \caption{Accuracy of estimated filters for different methods.
             We compare a Wiener filter, adaptive filtering, the IR-MLP trained with a conventional $ \ell_2 $-loss, and the IR-MLP with the noise robust loss from Section~\ref{sec:learning}.
             The systems are evaluated for different signal-to-noise ratios, ranging from moderately noisy data (0dB SNR) to highly noisy data (-30dB SNR).
             We report the SDR between estimated and ground truth filter on the y-axis.
             Only the IR-MLP trained with the noise robust loss is able to estimate accurate filters under highly noisy conditions.
             }
    \label{fig:learning_with_noise}
\end{figure*}

Practical applications typically include noise, thus a more accurate description of the system than the one in Equation~\eqref{eq:lti} is
\begin{align}
    y(t) = s(t) * h(t) + n(t),
\end{align}
where $ n(t) $ is an additive noise term.
Note that both $ h(t) $ and $ n(t) $ can potentially be position dependent.
To simplify notation, we drop the positional indices.
In traditional approaches such as Wiener filtering, noise is typically assumed to be stationary and to have known spectral characteristics.
While stationarity is a reasonable assumption, spectral characteristics of noise are generally unknown in practice.
Here, we propose a technique to learn the ideal impulse responses in noisy systems with stationary noise but unknown spectral noise characteristics.
This is achieved by a learned noise model, see Figure~\ref{fig:loss}.

Given a source signal $ s(t) $ and a measured, noisy target signal $ \hat y(t) $, the ideal IR and noise estimates $ h(t) $ and $ n(t) $ minimize
\begin{align}
    \mathcal{L} = \sum_t \Big(\hat y(t) - \big[s(t)*h(t) + n(t)\big] \Big)^2.
    \label{eq:loss_basic}
\end{align}
It is generally impossible to optimize this function because the exact form of $ n(t) $ is uncorrelated to the model's input.
With the assumption that noise is stationary, however, we can bypass this problem.
We define the residual $ r(t) := \hat y(t) - s(t)*h(t) $ and apply Parseval's theorem such that we obtain
\begin{align}
    \mathcal{L} &= \sum_t \big(r(t) - n(t)\big)^2 \nonumber 
                    = \sum_\omega |R(\omega) - N(\omega)|^2 \nonumber \\
                = &\sum_\omega |R(\omega)|^2 + |N(\omega)|^2 - 2|R(\omega)||N(\omega)|\cos(\phi_R - \phi_N),
    \label{eq:exact_noise_loss}
\end{align}
where $ \phi_R $ and $ \phi_N $ are the phase of the residual and noise, respectively.
We assume that the stationary noise best explains the residual between the measured signal $ \hat y(t) $ and the result of the convolution $ s(t) * h(t) $.
This is the case if $ \phi_R $ and $ \phi_N $ are chosen such that Equation~\eqref{eq:exact_noise_loss} is minimal, which is the case for $ \phi_R = \phi_N $.
Then,
\begin{align}
    \mathcal{L} &= \sum_\omega |R(\omega)|^2 + |N(\omega)|^2 - 2|R(\omega)||N(\omega)| \nonumber \\
                %&= \sum_\omega \Big(|R(\omega)| - |N(\omega)| \Big)^2 \nonumber \\
                &= \sum_\omega \Big(|\hat Y(\omega) - S(\omega)H(\omega)| - |N(\omega)| \Big)^2.
    \label{eq:noise_loss}
\end{align}
In contrast to Equation~\eqref{eq:loss_basic}, this function can be optimized since it no longer depends on $ n(t) $ but only on the noise amplitude spectrum $ |N(\omega)| $.
The latter is either a learnable constant if we assume position independent noise, or a learnable function of $ (x, y, z) $ if we assume position dependent noise.
Hence, given a model that predicts both $ h(t) $ and $ |N(\omega)| $ as in Figure~\ref{fig:loss}, the weights of the IR-MLP and the Noise MLP can simultaneously be learnt by optimization of Equation~\eqref{eq:noise_loss}.
%
%In Equation~\eqref{eq:noise_loss}, only the filter $ h(t) $ and the amplitude spectrum $ |N(\omega)| $ of the stationary noise are unknown.
%
%As before, $ h(t) $ is predicted by an MLP.
%
%To predict the noise amplitude spectrum $ |N(\omega)| $, we employ a similar MLP.
%
%The weights of the IR-MLP and the Noise MLP are simultaneously learnt by backpropagating the loss from Equation~\eqref{eq:noise_loss} through both networks, \cf Figure~\ref{fig:loss}.
%
In this formulation, stationarity is the only constraint on noise, knowledge of any spectral characteristics of the noise is not required.
%
%Moreover, noise can be position dependent, \ie the spectral characteristics of the stationary noise can depend on the spatial location of the signal measurement.
%
%\Alex{We'll get some push back on this. You also have access to the source signal and traditional adaptive filtering algorithms also remove noise based on viewing the data (FxLMS in particular). Sine sweeps are just a particular outlier example because of the delta freq over time. I don't think we have the results to justify broadly claiming improvement over adaptive filtering methods.}Note the advantage of this formulation with a learnable noise amplitude spectrum: in contrast to traditional signal processing models, spectral characteristics of noise can be learnt directly from data, which allows to handle any kind of stationary noise without prior knowledge or assumptions on its characteristics.

%%%%%%%%%%%%%%%%%%%%%%%%%%%%%%%%%%%%%%%%%%%%%%%%%%%%%%%%%%%%%%%%%%%%%%%%%%%%%%%%%%%%%%%%%%%%%%%%%%%%%%%%%%%%%%%%%%%%%%%%%%%%%%%%%%%%%%%%%%%%%%%%%%%%%%%%%%%%
%%% EVALUTATION
%%%%%%%%%%%%%%%%%%%%%%%%%%%%%%%%%%%%%%%%%%%%%%%%%%%%%%%%%%%%%%%%%%%%%%%%%%%%%%%%%%%%%%%%%%%%%%%%%%%%%%%%%%%%%%%%%%%%%%%%%%%%%%%%%%%%%%%%%%%%%%%%%%%%%%%%%%%%

\section{Evaluation}
\label{sec:evaluation}

\begin{table*}[tb]
    \footnotesize
    \centering
    \begin{tabularx}{1.0\textwidth}{lXrXrXrXr}
        \toprule
        \textbf{Architecture}       & & \multicolumn{5}{c}{\textbf{Efficiency}}                                    & & \textbf{Quality} \\
                                        \cmidrule(lr){3-7}                                                             \cmidrule(lr){9-9}
                                    & & parameters        & & compression       & & IR generation time             & & SDR of predicted IRs \\
        \midrule
        layers=6, hidden size=32    & & $ 9.1k $          & & $ 99.88\% $       & & $ 0.646 $ ms $ \pm 0.164 $ ms  & & $ 11.993 $ dB \\
        layers=6, hidden size=64    & & $ 30.5k $         & & $ 99.61\% $       & & $ 0.985 $ ms $ \pm 0.247 $ ms  & & $ 15.361 $ dB \\
        layers=6, hidden size=128   & & $ 101.2k $        & & $ 98.58\% $       & & $ 1.551 $ ms $ \pm 0.313 $ ms  & & $ 19.172 $ dB \\
        layers=6, hidden size=256   & & $ 417.0k $        & & $ 94.64\% $       & & $ 3.890 $ ms $ \pm 0.400 $ ms  & & $ 21.472 $ dB \\
        layers=6, hidden size=512   & & $ 1.62M $         & & $ 79.16\% $       & & $ 12.975 $ ms $ \pm 0.822 $ ms & & $ 24.437 $ dB \\
        \bottomrule
    \end{tabularx}
    \vspace{-0.3cm}
    \caption{A comparison of the memory- and computational efficiency versus the quality of predicted IRs for IR-MLPs of different sizes.
             IR-MLPs allow for a significant compression over naively storing the raw IRs, which would require $7.7$M floats, while still being able to recover the parameterized IRs at high quality.}
    \label{tab:efficiency}
    \vspace{-0.3cm}
\end{table*}

%\subsection{Experimental Setup}

%We empirically evaluation of our approach along the axes of robust IR estimation under noisy conditions, accurate IR interpolation, and efficient parameterization.
%
To show the effectiveness of our approach, we evaluate on synthetic data generated from measured HRIRs, such that ground truth IRs are known and we have full control over the noise.
%
%In Section~\ref{sec:real_data}, we also demonstrate effectiveness of our approach on in-the-wild data.
%
%HRIRs are a familiy of IRs that are dependent on spatial positions and are typically measured densely at several thousand positions~\cite{armstrong2018perceptual}.
%
%They therefore fit well into the proposed framework and are a suitable example to evaluate the effectiveness of our approach.

\textit{Dataset.}
We use an in-house dataset of 9,720 HRIRs measured on a sphere around a listener in an anechoic chamber, each of which is a two-channel FIR with $ 400 $ taps per channel.
We synthetically generate the observed (target) signals $ \hat y(t) $ by convolving a logarithmic sine sweep (the source signal $ s(t) $) with each of the measured IRs and add different kinds of noise.
%Sine sweeps are chosen for their ubiquitous usage as test signals for acoustic system identification.
%In addition, traditional adapative filtering algorithms typically require broadband noise excitation.
%
In Section~\ref{sec:real_data}, we show that our approach works on noisy real-world data as well.
All audio data is sampled at $ 48 $kHz.

\textit{Network Architectures.}
The IR-MLP, which predicts the IRs, consumes the spatio-temporal coordinates $ (x, y, z, t) $ and maps them onto a higher dimensional space using Fourier features from Equation~\eqref{eq:fourier_features} with $ L = 10 $, followed by six fully connected layers with 512 hidden units each.
If we assume position independent noise in the signal, the noise model is a simple learnable vector with $ 1,024 $ components representing the static noise amplitude spectrum.
In case of position dependent noise, the noise model consumes the spatial coordinates as input and produces a position dependent $ 1,024 $ component noise amplitude spectrum as illustrated in Figure~\ref{fig:loss}.
In this case, we use the same architecture as for the IR-MLP with four instead of six layers.
%
%We will release the source code upon acceptance of the paper.

\textit{Evaluation Protocol.}
We report the signal-to-distortion ratio (SDR) between ground truth IRs and estimated IRs,
\begin{align}
    \mathrm{SDR}(h_\text{true}, h_\text{est}) = 10 \log_{10} \Big( \frac{\|h_\text{true}\|^2}{\|h_\text{true} - h_\text{est}\|^2} \Big).
\end{align}

\subsection{Learning from noisy data}

We start with an evaluation of the robustness of our approach on noisy data.
We investigate two scenarios: learning from data with position independent noise and with position dependent noise.

\textit{Position independent noise.}
To generate target signals $ \hat y(t) $ with position independent noise, we randomly sample amplitude spectral values and scale the resulting noise spectrogram such that the signal-to-noise ratio (SNR) is at a predefined level and add random phase to the signal.
%for the frequency bands 3-6kHz, 9-12kHz, 15-18kHz, and 21-24kHz. The frequency bands in between remain noise-free. \Dejan{Is there a reason for this? It looks as an odd choice and could raise the question weather the method works for a broadband noise.}
%We scale the resulting noise spectrogram such that the signal-to-noise ratio (SNR) is at a predefined level and add random phase to the signal.
%
The noisy target data is then obtained by adding the generated noise to the result of the convolution of the sine sweep and the ground truth IR.

\textit{Position dependent noise.}
Position dependent noise, or noise with spectral charachteristics that change based on the spatial position of sound source or listener, is generated similarly to position independent noise.
We define two noise frequency bands of width 3kHz which are moved along the spectrum depending on the input positions, \ie, the input position defines the frequency band of the noise.
As before, we sample random phase and add the position dependent noise to the result of the convolution of the sine sweep and the ground truth IR.

We compare how accurately our IR-MLP learns the impulse responses when trained with a simple $ \ell_2 $-loss on the raw waveforms and when trained with the noise-robust loss proposed in Section~\ref{sec:learning}.

\textit{Baselines.}
As a baseline, we report the quality of IR estimation using adaptive filtering and a Wiener filter.
Since the convergence of adaptive filters requires excitation signals with broadband energy and minimal autocorrelation, we replaced the sine sweep excitation used for the other methods with white noise for the adaptive filtering baseline.
%
%While more sophisticated adaptive filtering techniques exist, many require white noise excitation to converge and, as such, could not be investigated given our data set. 
%
%The frequency domain Wiener filter can be formulated as
%\begin{align}
%    W(\omega) = \frac{P_{s}(\omega)}{P_{s}(\omega) + P_{n}(\omega)},
%    \label{eq:freqWiener}
%\end{align}
%where $W(\omega)$ is the frequency response of the Wiener filter, $P_{s}(\omega)$ is the power spectrum of the source signal, and $P_n(\omega)$ is the power spectrum of the noise. The Wiener filter is required to remove the observed noise, without also altering the impulse response of the system under observation. This is done by deriving an estimated noisy impulse response,
%
The Wiener filter requires knowledge about the spectral characteristics of noise $ |N(\omega)| $.
%, which are usually unknown in practical applications.
%
In order to estimate the characteristics of noise, first a noisy impulse response
\begin{align}
    \hat{h}_{n}(t) = \mathfrak{F}^{-1}\left(\frac{\hat Y(\omega)}{S(\omega)}\right) = \mathfrak{F}^{-1}\left(H(\omega) + \frac{N(\omega)}{S(\omega)}\right)
    \label{eq:noisyIR}
\end{align}
is estimated from the source signal $ s(t) $ and the observed (target) signal $ \hat y(t) $.
Per~\cite{lundeby1995rt}, the final ten percent of the impulse response is assumed to be noise-dominated.
While the spectrum of $\hat{h}_{n}$ does not contain the true spectrum $ |N(\omega)| $ of the additive noise in the system, we obtain an estimate for $ |N(\omega)| $ by multiplication of the spectrum of the noise-dominated portion of $\hat{h}_{n}$ with $ S(\omega) $.
The resulting Wiener filter is then applied to the observed signal $ \hat y(t) $ to recover the estimate of the denoised impulse response of the system under observation.
%\Dejan{Would it make sense to include results for ideal Wiener filter that instead of an estimate has the true noise spectrum?}

%
\textit{Results.}
For both, position independent noise (Figure~\ref{fig:learning_with_noise}a) and position dependent noise (Figure~\ref{fig:learning_with_noise}b), adaptive filtering, the Wiener filter, and the IR-MLP trained with a conventional $ \ell_2 $-loss fail to estimate the filters accurately in noisy scenarios with a negative SNR.
Training the IR-MLP with the noise-robust loss from Section~\ref{sec:learning}, on the contrary, allows for accurate filter estimation even in highly noisy scenarios with an SNR as low as -30dB.

\subsection{Interpolation: Predicting Unseen Impulse Responses}
\label{sec:interpolation}

\begin{figure}[tb]
    \centering
    \includegraphics[scale=0.9]{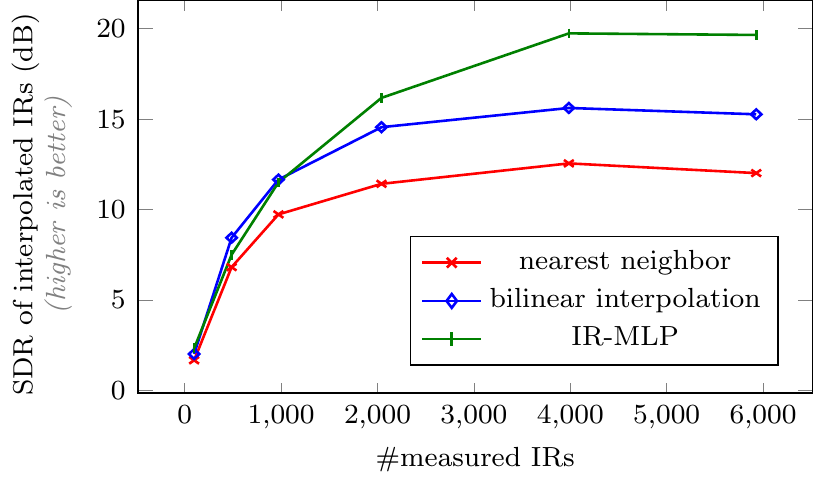}
    \vspace{-0.3cm}
    \caption{Quality of interpolated IRs at new (unmeasured) positions from a sparse set of measured IRs.
    As the number of available measurements increases (x-axis), the quality of interpolated IRs also increases.
    The IR-MLP achieves much higher quality results compared to a nearest-neighbor baseline and bilinear interpolation.}
    \label{fig:interpolation}
    \vspace{-0.3cm}
\end{figure}

%Interpolation of impulse responses at unmeasured spatial positions is a challenging problem.
%
%Traditional interpolation approaches like bilinear or barycentric interpolation do not always yield convincing results.
%
%More sophisticated interpolation requires domain-specific research and different approaches emerged for different families of filters such as RIRs~\cite{das2021room} or HRIRs~\cite{gamper2013hrtfinterp,gustavo2009hrtf}.
%
%Neural networks, on the contrary, are excellent interpolation machines.
%
%Since the IR-MLP receives spatial coordinates as input and predicts the filter from these coordinates, it can naturally generate filters at unseen spatial positions.

In this section, we evaluate the quality of the IR-MLP to predict IRs at positions that are not contained in the set of measured impulse responses.
From the entirety of 9,720 measured IRs in the above described HRIR database, we train an IR-MLP using data generated from a limited number of the measured IRs and evaluate how accurately the network can predict the impulse responses at positions that were not part of the training data.
The results are shown in Figure~\ref{fig:interpolation}.
We compare to a simple baseline where the nearest measured IR from the training set is returned as approximation to the unseen spatial position (nearest neighbor) and to bilinear interpolation.
The number of available impulse responses to generate the training data is gradually increased from 100 to 6,000.
Not surprisingly, all methods perform poorly when less than 1,000 measurements are available.
However, traditional interpolation methods such as bilinear interpolation saturate at around 15dB SDR, while the IR-MLP produces much higher quality interpolations with an SDR up to 20dB if the spatial density of measured impulse responses is high enough.

\subsection{Efficient Parameterization}

Besides the previously outlined advantages of estimating and parameterizing IRs with neural networks (robustness to noise, native interpolation to unseen positions), IR-MLPs provide a memory efficient way to store IRs while being able to recover them with low computational effort.
This is an important property for on-device applications where memory and compute are scarce resources.
We demonstrate the trade-offs between efficiency and quality of IR-MLPs in Table~\ref{tab:efficiency} with an example of the HRIR database with 9,720 measured IRs.
Storing all measured filters directly requires as much as $ 7.7 $M floats.
Parameterizing them in a IR-MLP, on the other hand, allows for almost lossless IR recovery ($\approx 20$dB SDR) with as few as 100k floats, which is a compression of more than $ 98\% $ compared to naively storing all measured filters.
Restoring the IRs requires a forward pass through the network.
With the 100k parameter model (6 hidden layers, 128 hidden units per layer), this requires only $ 1.5 $ms on a single CPU core on a Macbook Pro.
Different neural network sizes equip this approach with great flexibility: in applications where memory efficiency is paramount, the network size can easily be reduced until the requirements are met.
In applications where quality is paramount, on the other hand, larger networks allow for higher accuracy of predicted IRs at the cost of memory and compute.

\subsection{Learning from in-the-wild data}
\label{sec:real_data}

\begin{table}[tb]
    \footnotesize
    \centering
    \begin{tabularx}{0.48\textwidth}{Xrrr|r}
        \toprule
        \multicolumn{2}{r}{$ \ell_2 $ loss ($\times 10^3) $}                                & power                & phase                  & latency \\
        \midrule
        DSP baseline~\cite{richard2021binaural}     & $ 0.485 $                             & $ 0.058 $            & $ 1.388 $              & $ 25.0 $ ms\\
        neural renderer~\cite{richard2021binaural}  & $ \mathbf{0.167} $                    & $ 0.048 $            & $ \mathbf{0.807} $     & $ ^\dagger32.8 $ ms\\
        IR-MLP \textit{(ours)}                      & $ 0.236 $                             & $ \mathbf{0.042} $   & $ 0.933 $              & $ \mathbf{13.1} $ ms\\
        \bottomrule
    \end{tabularx}
    \caption{Performance of rendering binaural audio where the IRs are learned from real-world speech data.
             IR-MLPs clearly outperform a traditional DSP baseline and are almost as strong as the neural renderer from~\cite{richard2021binaural}.
             IR-MLPs can be run at much lower latency than the approach from~\cite{richard2021binaural}, which even requires a GPU (indicated by $\dagger$).}
    \label{tab:real_world_data}
    \vspace{-0.5cm}
\end{table}

We demonstrate that our approach can learn accurate impulse responses from noisy in-the-wild data.
Therefore, we use a 2h dataset of binaural speech recordings and tracked source and listener positions~\cite{richard2021binaural}.
Since the ground truth binaural impulse responses are unknown for such a real-world dataset, we follow the evaluation protocol from~\cite{richard2021binaural} and report $ \ell_2 $-loss on the binauralized speech data and on the power spectrum as well as angular error on the phase spectrum.
For the IR-MLP, we use six layers with 512 hidden units per layer as in the experiments above.
The IR-MLP successfully learns binaural impulse responses from the in-the-wild data, resulting in significantly better performance than a traditional signal processing baseline has, see Table~\ref{tab:real_world_data}.
Compared to the neural renderer from~\cite{richard2021binaural}, our approach performs slightly worse, particularly due to a higher phase angular error.
Note, however, that~\cite{richard2021binaural} is heavily optimized for binaural rendering and includes physical priors such as time warping, whereas our approach is a generic neural network without explicit assumptions and domain-specific components.
Also note that~\cite{richard2021binaural} requires a GPU to run at 32ms latency, while our approach runs with only 13ms latency on a single CPU core.
The improved latency compared to the DSP baseline (13ms vs.\ 25ms) can be explained by a lower number of samples that need to be buffered for our approach:
the network can produce IRs at dense spatial positions that change smoothly over time, which allows it to produce smooth audio outputs even with small frame sizes and without an overlap-add operation, saving compute and reducing latency.

%%%%%%%%%%%%%%%%%%%%%%%%%%%%%%%%%%%%%%%%%%%%%%%%%%%%%%%%%%%%%%%%%%%%%%%%%%%%%%%%%%%%%%%%%%%%%%%%%%%%%%%%%%%%%%%%%%%%%%%%%%%%%%%%%%%%%%%%%%%%%%%%%%%%%%%%%%%%
%%% CONCLUSION
%%%%%%%%%%%%%%%%%%%%%%%%%%%%%%%%%%%%%%%%%%%%%%%%%%%%%%%%%%%%%%%%%%%%%%%%%%%%%%%%%%%%%%%%%%%%%%%%%%%%%%%%%%%%%%%%%%%%%%%%%%%%%%%%%%%%%%%%%%%%%%%%%%%%%%%%%%%%

\section{Conclusion}
\label{sec:conclusion}

We proposed a framework that can effectively estimate and interpolate IRs using deep neural networks which leverage recent advances in neural representation learning.
Our approach proves to be robust to low signal-to-noise ratios in the observed signals and allows to handle in-the-wild data.
The latter is a particularly important advancement as it allows to estimate IRs directly from user-collected data rather than from idealized lab recordings that require complex and costly equipment.
In its current formulation, the framework is generic and directly applicable to a wide variety of IR estimation and interpolation problems.
%
%In future works, we aim to further explore the proposed framework by combining it with ideas from traditional audio and speech modeling applications, \eg in room acoustics estimation and simulation, or speech enhancement.
%However, we consider such downstream tasks beyond the scope of this paper and leave them for future work.

\vfill\pagebreak

\bibliographystyle{IEEEbib}
\bibliography{references}

\begin{thebibliography}{10}

\bibitem{lin2006bayesian}
Yuanqing Lin and Daniel~D Lee,
\newblock ``Bayesian regularization and nonnegative deconvolution for room
  impulse response estimation,''
\newblock {\em IEEE Transactions on Signal Processing}, 2006.

\bibitem{szoke2019building}
Igor Sz{\"o}ke, Miroslav Sk{\'a}cel, Ladislav Mo{\v{s}}ner, Jakub Paliesek, and
  Jan {\v{C}}ernock{\`y},
\newblock ``Building and evaluation of a real room impulse response dataset,''
\newblock {\em IEEE Journal of Selected Topics in Signal Processing}, 2019.

\bibitem{xie2007head}
BoSun Xie, XiaoLi Zhong, Dan Rao, and ZhiQiang Liang,
\newblock ``Head-related transfer function database and its analyses,''
\newblock {\em Science in China Series G: Physics, Mechanics and Astronomy},
  2007.

\bibitem{dinuzzo2015kernels}
Francesco Dinuzzo,
\newblock ``Kernels for linear time invariant system identification,''
\newblock {\em SIAM Journal on Control and Optimization}, 2015.

\bibitem{wiener1949filter}
Norbert Wiener,
\newblock {\em Extrapolation, Interpolation, and Smoothing of Stationary Time
  Series with Engineering Applications},
\newblock The MIT Press, 1949.

\bibitem{widrow1975lms}
Bernard Widrow, John~R Glover, John~M McCool, John Kaunitz, Charles~S Williams,
  Robert~H Hearn, James~R Zeidler, Eugene Dong~Jr., and Robert~C Goodlin,
\newblock ``Adaptive noise cancelling: Principles and applications,''
\newblock {\em Proceedings of the {IEEE}}, 1975.

\bibitem{liu2010adaptivefiltering}
Weifeng Liu, Jose~C Principe, and Simon Haykin,
\newblock {\em Kernel Adaptive Filtering: A Comprehensive Introduction},
\newblock Wiley, 2010.

\bibitem{antonello2017room}
Niccolo Antonello, Enzo De~Sena, Marc Moonen, Patrick~A Naylor, and Toon
  Van~Waterschoot,
\newblock ``Room impulse response interpolation using a sparse spatio-temporal
  representation of the sound field,''
\newblock {\em IEEE/ACM Transactions on Audio, Speech, and Language
  Processing}, 2017.

\bibitem{ramos2013parametric}
German Ramos and Maximo Cobos,
\newblock ``Parametric head-related transfer function modeling and
  interpolation for cost-efficient binaural sound applications,''
\newblock {\em The Journal of the Acoustical Society of America}, 2013.

\bibitem{gamper2013hrtfinterp}
Hannes Gamper,
\newblock ``Head-related transfer function interpolation in azimuth, elevation,
  and distance,''
\newblock {\em The Journal of the Acoustical Society of America}, 2013.

\bibitem{das2021room}
Orchisama Das, Paul Calamia, and Sebastia V~Amengual Gari,
\newblock ``Room impulse response interpolation from a sparse set of
  measurements using a modal architecture,''
\newblock in {\em IEEE Int. Conf. on Acoustics, Speech and Signal Processing},
  2021.

\bibitem{sitzmann2019siren}
Vincent Sitzmann, Julien~N.P. Martel, Alexander~W. Bergman, David~B. Lindell,
  and Gordon Wetzstein,
\newblock ``Implicit neural representations with periodic activation
  functions,''
\newblock in {\em Advances in Neural Information Processing Systems}, 2020.

\bibitem{tancik2020fourierfeatures}
Matthew Tancik, Pratul~P. Srinivasan, Ben Mildenhall, Sara Fridovich-Keil,
  Nithin Raghavan, Utkarsh Singhal, Ravi Ramamoorthi, Jonathan~T. Barron, and
  Ren Ng,
\newblock ``Fourier features let networks learn high frequency functions in low
  dimensional domains,''
\newblock {\em Advances in Neural Information Processing Systems}, 2020.

\bibitem{mildenhall2020nerf}
Ben Mildenhall, Pratul~P. Srinivasan, Matthew Tancik, Jonathan~T. Barron, Ravi
  Ramamoorthi, and Ren Ng,
\newblock ``Nerf: Representing scenes as neural radiance fields for view
  synthesis,''
\newblock in {\em European Conf. on Computer Vision}, 2020.

\bibitem{vandenoord2016wavenet}
A{\"a}ron Van Den~Oord, Sander Dieleman, Heiga Zen, Karen Simonyan, Oriol
  Vinyals, Alex Graves, Nal Kalchbrenner, Andrew~W. Senior, and Koray
  Kavukcuoglu,
\newblock ``Wavenet: A generative model for raw audio,''
\newblock in {\em ISCA Speech Synthesis Workshop}, 2016.

\bibitem{fu2017raw}
Szu-Wei Fu, Yu~Tsao, Xugang Lu, and Hisashi Kawai,
\newblock ``Raw waveform-based speech enhancement by fully convolutional
  networks,''
\newblock in {\em Asia-Pacific Signal and Information Processing Association
  Annual Summit and Conference}, 2017.

\bibitem{luo2018tasnet}
Yi~Luo and Nima Mesgarani,
\newblock ``Tasnet: time-domain audio separation network for real-time,
  single-channel speech separation,''
\newblock in {\em IEEE Int. Conf. on Acoustics, Speech and Signal Processing},
  2018.

\bibitem{donahue2019wavegan}
Chris Donahue, Julian McAuley, and Miller Puckette,
\newblock ``Adversarial audio synthesis,''
\newblock in {\em Int. Conf. on Learning Representations}, 2019.

\bibitem{richard2014mnsgd}
Simon Wiesler, Alexander Richard, Ralf Schluter, and Hermann Ney,
\newblock ``Mean-normalized stochastic gradient for large-scale deep
  learning,''
\newblock in {\em IEEE Int. Conf. on Acoustics, Speech and Signal Processing},
  2014.

\bibitem{pepe2020designing}
Giovanni Pepe, Leonardo Gabrielli, Stefano Squartini, and Luca Cattani,
\newblock ``Designing audio equalization filters by deep neural networks,''
\newblock {\em Applied Sciences}, 2020.

\bibitem{nossier2020comparative}
Soha~A Nossier, Julie Wall, Mansour Moniri, Cornelius Glackin, and Nigel
  Cannings,
\newblock ``A comparative study of time and frequency domain approaches to deep
  learning based speech enhancement,''
\newblock in {\em Int. Joint Conf. on Neural Networks}, 2020.

\bibitem{richard2021binaural}
Alexander Richard, Dejan Markovic, Israel~D Gebru, Steven Krenn, Gladstone
  Butler, Fernando de~la Torre, and Yaser Sheikh,
\newblock ``Neural synthesis of binaural speech from mono audio,''
\newblock in {\em Int. Conf. on Learning Representations}, 2021.

\bibitem{gebru2021implicit}
Israel~D Gebru, Dejan Markovic, Alexander Richard, Steven Krenn, Gladstone
  Butler, Fernando de~la Torre, and Yaser Sheikh,
\newblock ``Implicit hrtf modeling using temporal convolutional networks,''
\newblock in {\em IEEE Int. Conf. on Acoustics, Speech and Signal Processing},
  2021.

\bibitem{steinmetz2021filtered}
Christian~J Steinmetz, Vamsi~Krishna Ithapu, and Paul Calamia,
\newblock ``Filtered noise shaping for time domain room impulse response
  estimation from reverberant speech,''
\newblock {\em arXiv preprint arXiv:2107.07503}, 2021.

\bibitem{ratnarajah2020ir}
Anton Ratnarajah, Zhenyu Tang, and Dinesh Manocha,
\newblock ``Ir-gan: Room impulse response generator for far-field speech
  recognition,''
\newblock {\em Interspeech}, 2021.

\bibitem{gamper2018blind}
Hannes Gamper and Ivan~J Tashev,
\newblock ``Blind reverberation time estimation using a convolutional neural
  network,''
\newblock in {\em 2018 16th International Workshop on Acoustic Signal
  Enhancement (IWAENC)}, 2018.

\bibitem{bryan2020impulse}
Nicholas~J Bryan,
\newblock ``Impulse response data augmentation and deep neural networks for
  blind room acoustic parameter estimation,''
\newblock in {\em IEEE Int. Conf. on Acoustics, Speech and Signal Processing},
  2020.

\bibitem{yu2020room}
Wangyang Yu and W~Bastiaan Kleijn,
\newblock ``Room acoustical parameter estimation from room impulse responses
  using deep neural networks,''
\newblock {\em IEEE/ACM Transactions on Audio, Speech, and Language
  Processing}, 2020.

\bibitem{lundeby1995rt}
Anders Lundeby, H.~Bietz, T.~Vigran, and Michael Vorlander,
\newblock ``Uncertainties of measurements in room acoustics,''
\newblock {\em Acta Acustica united with Acustica}, 1995.

\end{thebibliography}

\end{document}